\documentstyle{article}

\def\stackunder#1#2{\mathrel{\mathop{#2}\limits_{#1}}}

\def\QATOPD#1#2#3#4{{#3 \atopwithdelims#1#2 #4}}%

\begin{document}

\title{Internal Time Peculiarities as a Cause of Bifurcations Arising in
Classical Trajectory Problem and Quantum Chaos Creation in Three-Body System }
\author{Bogdanov A.V.$^1$, Gevorkyan A.S.$^2$, Grigoryan A.G.$^3$, Matveev S.A.}
\date{Institute for High-Performance Computing and Data Bases\\
P/O Box 71, 194291, St-Petersburg, Russia\\
$^1$bogdanov@hm.csa.ru, $^2$ashot@fn.csa.ru, $^3$ara@fn.csa.ru}
\maketitle

\begin{abstract}
A new formulation of the theory of quantum mechanical multichannel scattering
for three-body collinear systems is proposed. It is shown, that in this simple
case the principle of quantum determinism in the general case breaks down and
we have a micro-irreversible quantum mechanics. The first principle
calculations of the quantum chaos (wave chaos) were pursued on the example of
an elementary chemical reaction $Li+\left( FH\right) \rightarrow \left(
LiFH\right) ^{*}\rightarrow \left( LiF\right) +H$.
\end{abstract}

%%%%%%%%%%%%%%%%%%%%%%%%%%%%%%%%%%%%%%%%%%%%%%%%%%%%%%%%%%%%%%%%%%%%%%%%%%
%%%%%%%%%%%%%%%%%%%%%%%%%%%%%%%%%%%%%%%%%%%%%%%%%%%%%%%%%%%%%%%%%%%%%%%%%%
%%                                                                      %%
%%                                                                      %%
%%   Глава 1. Введение                                                  %%
%%                                                                      %%
%%                                                                      %%
%%%%%%%%%%%%%%%%%%%%%%%%%%%%%%%%%%%%%%%%%%%%%%%%%%%%%%%%%%%%%%%%%%%%%%%%%%
%%%%%%%%%%%%%%%%%%%%%%%%%%%%%%%%%%%%%%%%%%%%%%%%%%%%%%%%%%%%%%%%%%%%%%%%%%

\section{Introduction}

At the early stage of quantum mechanics development A. Einstein had asked 
the question that attracted close attention several decades later
\cite{Einstein}. The question was: what will the classic chaotic system
become in terms of quantum mechanics. He has particularly set apart the
three-body system.

In an effort to formulate and obtain the solution for the problem of quantum
chaos M. Gutzwiller has tentatively subdivided all the existing knowledge
in physics into three areas \cite{Gutzwiller}:

1) regular classical mechanics (area $R$);

2) classical chaotic systems or dynamical systems of Poincar\'{e} ($P$ area);

3) regular quantum mechanics ($Q$ area).

The above areas are connected by certain conditions. Thus, the Bohr's
correspondence principle works between $R$ and $Q$ areas, transferring the
quantum mechanics into classical Newton's mechanics within the limit $\hbar
\rightarrow 0$. Areas $R$ and $P$ are connected by Kolmogorov's - Arnold's -
Moser's theorem (KAM). In spite of well known work by F. Nelson \cite{Nelson},
which allows to describe $Q$-systems with the help of $P$-systems in the
thermodynamical limit under certain circumstances, the general principle
connecting $P$ and $Q$ is not determined yet. Assuming the existence of a
fourth area - quantum chaos area $Q_{ch}$, M. Gutzwiller adds that the
"quantum chaos" conception is rather a puzzle than a well formulated
problem. It is evident that the task formulated correctly in $Q_{ch}$ area
is a most general one and under specific conditions must transform into
the above-mentioned limiting areas.

The problem of quantum chaos was studied by the authors taking as an example
the quantum mechanical multichannel scattering in collinear three-body system
\cite{Bogdanov}-\cite{Gevorkyan}. It was shown than this task can be
transformed into a problem of anharmonic oscillator with non-trivial time
(internal time), which in the general case can have a chaotic behaviour. In the
present work the study of the problem of quantum chaos is continued using
numerical calculations based on the example of an elementary chemical
reaction.

%%%%%%%%%%%%%%%%%%%%%%%%%%%%%%%%%%%%%%%%%%%%%%%%%%%%%%%%%%%%%%%%%%%%%%%%%%
%%%%%%%%%%%%%%%%%%%%%%%%%%%%%%%%%%%%%%%%%%%%%%%%%%%%%%%%%%%%%%%%%%%%%%%%%%
%%                                                                      %%
%%                                                                      %%
%%   Глава 2. Formulation of the problem                                %%
%%                                                                      %%
%%                                                                      %%
%%%%%%%%%%%%%%%%%%%%%%%%%%%%%%%%%%%%%%%%%%%%%%%%%%%%%%%%%%%%%%%%%%%%%%%%%%
%%%%%%%%%%%%%%%%%%%%%%%%%%%%%%%%%%%%%%%%%%%%%%%%%%%%%%%%%%%%%%%%%%%%%%%%%%

\section{Formulation of the problem}

The quantum multichannel scattering in the framework of collinear model is
formulated in such a way:

\begin{equation}
\label{eq2.1}
A+\left( BC\right) _n\rightarrow \left\{
\begin{array}{c}
A+\left( BC\right) _m\qquad \qquad \qquad \qquad  \\
\left( AB\right) _m+C\qquad \qquad \qquad \qquad  \\
A+B+C\qquad \qquad \qquad \qquad  \\
\left( ABC\right) ^{*}\rightarrow \left\{
\begin{array}{c}
A+\left( BC\right) _m \\
\left( AB\right) _m+C \\
A+B+C
\end{array}
\right.
\end{array}
\right.
\end{equation}

where $m$ and $n$ are the vibrational quantum numbers corresponding to
$\left( in\right) $ and $\left( out\right) $ scattering channels. As was
shown elsewhere \cite{Bogdanov}-\cite{Gevorkyan} one can formulate the
problem of quantum multichannel scattering (\ref{eq2.1}) as the motion of an
image point with reduced mass $\mu _0$ on the manifold $M$, that is a
lamination of the Lagrange surface within a local coordinate system moving on
$S_p$.  In our case there is a standard definition of the surface $S_p$

\begin{equation}
\label{eq2.2}
S_p=\left\{ x^1,x^2;2\mu _0\left( E-V\left( x^1,x^2\right) \right)
>0\right\} ,\qquad
\mu _0=\left\{ \frac{m_Am_Bm_C}{m_A+m_B+m_C}\right\} ^{1/2},
\end{equation}

where $m_A$, $m_B$, $m_C$ are the masses of corresponding particles, $E$ and
$V\left( x^1,x^2\right) $ are respectively the total energy and interaction
potential of the system. The metric on the surface $S_p$ in our case is
introduced in the following way

\begin{equation}
\label{eq2.3}
\begin{array}{c}
g_{ik}=P_0^2\left( x^1,x^2\right) \delta _{ik}, \\
\\
P_0^2\left( x^1,x^2\right) =2\mu _0\left( E-V\left( x^1,x^2\right) \right) .
\end{array}
\end{equation}

The motion of the local coordinate system is determined by the
projection of the image point motion over the extremal ray $\Im _{ext}$ of
the Lagrange manifold $S_p$. Note, that for scattering problem (\ref{eq2.1})
there are two extremal rays on a surface $S_p$: one connecting the $\left(
in\right) $ channel with the $\left( out\right) $ channel of particle
rearrangement and the other connecting the $\left( in\right) $ channel
with the $\left( out\right) $ channel, where all three particles are free.
From now on we shall study only the case of particles rearrangement at the
collision. Let us introduce curvilinear coordinates $\left( x^1,x^2\right) $
in Euclidean space $R^2$ along the projection of the rearrangement extremal
ray $\bar \Im _{ext}$ in a such way, that $x^1$ is changing along $\bar \Im
_{ext}$ and $x^2$ is changing in the orthogonal direction. In such a case the
trajectory of the image point is determined by the following system of second
order differential equations:

\begin{equation}
\label{eq2.4}
x_{;ss}^k+\QATOPD\{ \} {k}{ij}_{Sp}x_{;s}^ix_{;s}^j=0,\quad \left(
i,j,k=1,2\right)
\end{equation}

where $x^i_{;s}=\frac{dx^i}{ds}$ and $\QATOPD\{ \} {k}{ij}_{S_p}=\frac
12g^{kl}\left( \frac{\partial g_{lj}}{\partial x^i}+\frac{\partial
g_{il}}{\partial x^j}-\frac{\partial g_{ij}}{\partial x^l}\right) $.

The differential equations of second order (\ref{eq2.4}) with initial
conditions

\begin{equation}
\label{eq2.5}
x_{0}^{i}=x^{i}\left( -\infty \right) ,\qquad \dot x_{0}^{i}=x^{i}_{;t}\left(
-\infty \right)
\end{equation}

at any moment of time $t$ defines the only solutions $x^{i}\left( t\right) $
and $\dot x^{i}\left( t\right) $ - the geodesic trajectory and geodesic
velocity.

Now we pass to quantum description of rearrangement process. Let's note that
in quasiclassical limits such description equivalent to consideration of
geodesic trajectories flow on Lagrange surface $S_{p}$. It is convenient to
describe the flow in a local coordinate system given by  solution
$x^{1}\left( s\right) $ of the system (\ref{eq2.4}). The quantization
of corresponding trajectory flow should be carried out in indicated coordinate
system  on stratification $M$ of Lagrange surface $S_{p}$.

Taking into account the Schr\"odinger equation for arbitrary curvilinear
coordinate system $\left( x^{1},x^{2}\right) $ \cite{Dirak} and keeping in
mind aforesaid one can obtain the full wave function of the noted system in
moving local coordinates:

\begin{equation}
\label{eq2.6}
\left\{ \hbar ^2\Delta _{\left( x^1\left( s\right) ,x^2\right) }+P_0^2\left(
x^1\left( s\right) ,x^2\right) \right\} \Psi =0,
\end{equation}

with the operator $\Delta _{\left( x^1\left( s\right) ,x^2\right) }$ of the
form

\begin{equation}
\label{eq2.7}
\Delta _{\left( x^1\left( s\right) ,x^2\right) }=\gamma ^{-\frac 12}\left\{
\partial _{x^1\left( s\right) }\left[ \gamma ^{ij}\gamma ^{\frac 12}\partial
_{x^1\left( s\right) }\right] +\partial _{x^2}\left[ \gamma ^{ij}\gamma
^{\frac 12}\partial _{x^2}\right] \right\} ,\qquad
\partial _{x^{1}(s)}=\left. \partial \right/ \partial x^{i}\left( s\right) .
\end{equation}

The metric tensor of the manifold $M$ has the following form
\cite{Bogdanov}-\cite{Gevorkyan}:

\begin{equation}
\label{eq2.8}
\begin{array}{c}
\gamma _{11}=\left( 1+
\frac{\lambda \left( x^1\left( s\right) \right) }{\rho _1\left( x^1\left(
s\right) \right) }\right) ^2,\quad \gamma _{12}=\gamma _{21}=0, \\  \\
\gamma _{22}=\left( 1+\frac{x^2}{\rho _2\left( x^1\left( s\right) \right) }%
\right) ^2,\quad \gamma =\gamma _{11}\gamma _{22}>0,
\end{array}
\end{equation}

where $\lambda $ is de Broglie wave length on $\Im _{ext}$, $\rho _1$ and
$\rho _2$ being the principal curvatures of the surface $S_p$ in the point
$x^1\in \Im _{ext}$ in the directions of coordinates $x^1$ and $x^2$ which
are changed as

\begin{equation}
\label{eq2.9}
\rho _i^{-1}=\left. \frac{P_{0;x^i}\left( x^1\left( s\right) ,x^{2}\right) }%
{P_0\left( x^1\left( s\right) ,x^{2}\right) }\right| _{x^{2}=0},\quad
\lambda =\left. \frac \hbar {P_0\left( x^1\left( s\right) ,x^{2}\right) }%
\right| _{x^{2}=0},\quad
P_{0;x^i}=\frac{dP_{0}\left( x^1\left( s\right) ,x^2\right) }{dx^i}\qquad
i=1,2.
\end{equation}

Note, that the main difference of (\ref{eq2.6}) from Schr\"odinger equation
comes from the fact, that one of the independent coordinates $x^1\left(
s\right) $ is the solution of system of nonlinear differential equations and
as such, it is not a natural parameter of our system and can in certain
situations be a stochastic function.

Our purpose is to find a solution of equation (\ref{eq2.6}) that satisfy
the following asymptotic conditions for the total wave function of the system

$$
\stackunder{x^1\rightarrow -\infty }{\lim }\Psi ^{(+)}\left( x^1,x^2\right)
=\Psi _{in}\left( n;x^1,x^2\right) +\stackunder{m\neq n}{\sum }R_{mn}\Psi
_{in}\left( n;x^1,x^2\right) ,
$$

\begin{equation}
\label{eq2.10}
\stackunder{
\begin{array}{c}
^{x^1\rightarrow +\infty } \\
^{\left( s\rightarrow +\infty \right) }
\end{array}
}{\lim }\Psi ^{(+)}\left( x^1\left( s\right) ,x^2\right) =\stackunder{m}{\sum }%
S_{mn}R_{mn}\Psi _{out}\left( m;x^1,x^2\right) ,
\end{equation}

where the coefficients $R_{mn}$ and $S_{mn}$ are the excitation and
rearrangement amplitudes respectively.

%%%%%%%%%%%%%%%%%%%%%%%%%%%%%%%%%%%%%%%%%%%%%%%%%%%%%%%%%%%%%%%%%%%%%%%%%%
%%%%%%%%%%%%%%%%%%%%%%%%%%%%%%%%%%%%%%%%%%%%%%%%%%%%%%%%%%%%%%%%%%%%%%%%%%
%%                                                                      %%
%%                                                                      %%
%%   Глава 3. Reduction of the scattering problem to the problem...     %%
%%                                                                      %%
%%                                                                      %%
%%%%%%%%%%%%%%%%%%%%%%%%%%%%%%%%%%%%%%%%%%%%%%%%%%%%%%%%%%%%%%%%%%%%%%%%%%
%%%%%%%%%%%%%%%%%%%%%%%%%%%%%%%%%%%%%%%%%%%%%%%%%%%%%%%%%%%%%%%%%%%%%%%%%%

\section{Reduction of the scattering problem to the problem of quantum
harmonic oscillator with internal time}

Let us make a coordinate transformations in Eq.(\ref{eq2.6}):

\begin{equation}
\label{eq3.1}
\begin{array}{c}
\tau =\left( E_k^i\right) ^{-1}
\stackrel{x^1\left( s\right) }{\stackunder{0}{\int }}P\left( x^1,0\right)
\sqrt{\gamma _0}dx^1,
\qquad
z=\left( \hbar E_k^i\right) ^{-\frac 12}P\left( x^1,0\right) x^2,
\\  \\
\gamma _0=\left. \gamma \left( x^{1},x^{2}\right) \right| _{x^2=0},
\qquad
P\left( x^1,x^2\right) =\sqrt{2\mu _0\left[ E_k^i-V\left( x^1,x^2\right)
\right] },
\end{array}
\end{equation}

where $E_k^i$ is the kinetic energy of particle $A$ in the $\left(
in\right) $ channel and the function $P\left( x^1,x^2\right) $ is the
momentum of image point on the curve $\Im _{ext}$.

By expanding $P\left( x^1,x^2\right) $ in the coordinate $x^2$ up to the
second order we can reduce the scattering equation (\ref{eq2.6}) to the
problem of a quantum harmonic oscillator with variable frequency in an
external field, depending on internal time $\tau \left( x^1\left( s\right)
\right) $.  \cite{ZhTF}-\cite{TMF}. E.g., in the case of zero external field
the exact wave function of the system has the form

\begin{equation}
\label{eq3.2}
\Psi ^{(+)}\left( n;\tau \right) =\left[ \frac{\left( \Omega _{in}/\pi \right)
^{1/2}}{2^nn!\left| \xi \right| }\right] ^{\frac 12}
\exp \left\{ i\left[ \frac{E_k^i\tau }\hbar -i\left( n+\frac 12\right) \Omega
_{in}\stackrel{\tau }{\stackunder{-\infty }{\int }}\frac{d\tau ^{\prime
}}{\left| \xi \right| ^2}+\frac 12\dot \xi \xi ^{-1}z^2-\frac 12\dot
pp^{-1}z^2\right] \right\} H_n\left( \frac{\sqrt{\Omega _{in}}}{\left|
\xi \right| }z\right) ,
\end{equation}

$$
\dot \xi =d_\tau \xi ,\quad \dot p=d_\tau p,\quad p\left( x^1\left( s\right)
\right) =P\left( x^1\left( s\right) ,0\right) ,
$$

where the function $\xi \left( \tau \right) $ is the solution of the
classical oscillator equation

$$
\ddot \xi +\Omega ^2\left( \tau \right) \xi =0,
$$

\begin{equation}
\label{eq3.3}
\Omega ^2\left( \tau \right) =-\left( \frac{E_k^i}p\right) ^2\left\{ \frac
1{\rho _2^2}+\stackrel{2}{\stackunder{k=1}{\sum }}\left[ \frac{p_{;kk}}%
p+\left( \frac{p_{;k}}p\right) ^2\right] \right\} ,\qquad
p_{;k}=\frac{dp}{dx^k}
\end{equation}

with asymptotic conditions

\begin{equation}
\label{eq3.4}
\begin{array}{c}
\xi \left( \tau \right)
\stackunder{\tau \rightarrow -\infty }{\sim }\exp \left( i\Omega _{in}\tau
\right) , \\  \\
\xi \left( \tau \right) \stackunder{\tau \rightarrow +\infty }{\sim }C_1\exp
\left( i\Omega _{out}\tau \right) -C_2\exp \left( i\Omega _{out}\tau \right) .
\end{array}
\end{equation}

Note, that the internal time $\tau $ is directly determined by the solution
of $x^1\left( s\right) $ and therefore includes all peculiarities of $x^1$
solution.

The transition probability for that case is of the form
\cite{Bogdanov}-\cite{Gevorkyan}, \cite{ZhTF}-\cite{TMF}:

\begin{equation}
\label{eq3.5}
W_{mn}=\frac{n_{<}!}{n_{>}!}\sqrt{1-\rho }\left|
P_{\left( n_{>}+n_{<}\right) /2}^{\left( n_{>}-n_{<}\right) /2}\left(
1-\rho \right) \right| ^2, \qquad
\rho =\left| \frac{C_2}{C_1}\right| ^2,
\end{equation}

where $n_{<}=\min \left( m,n\right) $, $n_{>}=\max \left( m,n\right) $ and
$P_m^n$ being the Legendre polynomial.

%%%%%%%%%%%%%%%%%%%%%%%%%%%%%%%%%%%%%%%%%%%%%%%%%%%%%%%%%%%%%%%%%%%%%%%%%%
%%%%%%%%%%%%%%%%%%%%%%%%%%%%%%%%%%%%%%%%%%%%%%%%%%%%%%%%%%%%%%%%%%%%%%%%%%
%%                                                                      %%
%%                                                                      %%
%%   Глава 4. The study of the internal time dependence...              %%
%%                                                                      %%
%%                                                                      %%
%%%%%%%%%%%%%%%%%%%%%%%%%%%%%%%%%%%%%%%%%%%%%%%%%%%%%%%%%%%%%%%%%%%%%%%%%%
%%%%%%%%%%%%%%%%%%%%%%%%%%%%%%%%%%%%%%%%%%%%%%%%%%%%%%%%%%%%%%%%%%%%%%%%%%

\section{The study of the internal time dependence versus standard time -
the natural parameter of the problem}

Now we can turn to the proof of the chaos initiation possibility in the wave
function of the three-body system (\ref{eq3.2}) for the case of zero external
field, that we shall name as quantum chaos. Let's stress, that in this case
the transition probability dependence over classical trajectory features will
be nonregular too. It will be sufficient to show, that the solution
$x^1\left( s\right) $ under some initial conditions is unstable or chaotic.
For that purpose we studied indetail the behaviour of the image point
trajectories on Lagrange surface $S_p$ on example of elementary chemical
reaction $Li+\left( FH\right) _{n}\rightarrow \left( LiFH\right)
^{*}\rightarrow \left( LiF\right) _{m}+H$. The potential surface of this
reaction was reproduced using the quantum-mechanical calculations carried out
in work \cite{Surface}. It was shown, that if the beginning of geodesic
trajectory $x_{0}^{2}$ in the $\left( in\right) $ channel (i.e.
$x^{1}\rightarrow -\infty$) is fixed, then for the collision energy
$E_k^i\ge 1.4eV$ the behaviour of that trajectory is regular (Fig.1a,1b).
In this case the regular interchange of passing and reflection areas on the
energy scale is observed (Fig.2). The same observations are true for the case
when the energy $E_k^i$ is fixed and the value of $x^2_0$ changes.

Starting from $E_k^i\approx 1.4eV$ up to reaction threshold energy $1.1eV$
unstable behaviour of system motion can be seen in the flows of passing
and reflecting geodesic trajectories, that results in complete mixing in the
intermediate area at further reduction of $E^i_k$ energy and causes the
$\left( LiFH\right) ^{*}$ resonant complex generation.

It was shown by numerical calculations, that the largest Lyapunov exponents
are positive for all energy values. However, up to energy values of $1.4eV$
and greater its growth is very slow. And at energy interval from $1.4eV$ to
$1.1eV$ the quick increasing of largest Lyapunov exponents occurs. The last
fact shows the exponential divergence of trajectories. For the area of
collision energy $E^i_k$ mentioned the regular interchange of passing and
reflection fields is violated and the fields of unstable behaviour arises. The
numerical investigations of such fields shows theirs self-similarity relative
to scale transformation that characterizes them as fractal type objects
(see Fig.3).

%
%At the energy less than $E_k^i$ up to the
%reaction threshold energy $1.1eV$ there is a shuffling become apparent in
%behaviour of geodesic trajectories that result to $\left( LiFH\right) ^{*}$
%resonance complex initiation (Fig. 1c).  Let's remind, that after complex
%decay the image point trajectory might go to $\left( out\right) $ channel,
%but it might return to $\left( in\right) $ channel as well.
%
%The whole region of kinetic energies is
%splitted to regular subregions, and depending from which subregion trajectory
%starts it goes either to $\left( out\right) $ channel (Fig.2(a)) or reflects
%back in the $\left( in\right) $ channel (Fig.2(b)).
%
%With a further change of kinetic energy the image point trajectory in the
%interaction region starts orbiting, that corresponds to the creation of the
%resonance complex $\left( LiFH\right) ^{*}$, and after that leave the
%interaction region either to $\left( out\right) $ (Fig.2(c)) or return to
%$\left( in\right) $ channel. In such a case the image point trajectories
%diverge and this divergence is exponential, as can be seen from the study of
%the Lyapunov parameters on Fig.3.
%

Thus for those initial conditions the evolution in the correspondent classical
problem is chaotic and so the motion of the local coordinate system is chaotic
too. It is easy to see that in such situation the behaviour of $x^1\left(
s\right) $ is also chaotic and the same is true for internal time $\tau \left(
x^1\left( s\right) \right) $, that is the chronologization parameter of quantum
evolution in three-body system.

It can be shown, that chaotic behaviour of the internal time
$\tau \left( x^1\left( s\right) \right) $ is followed by
the stochastic behaviour of the model equation solution $\xi \left( \tau
\left( x^1\left( s\right) \right) \right) $. The same is true for the wave
function representation (\ref{eq3.2}) and transition probability (\ref{eq3.5}).
In such a way on the example of the simple multichannel scattering model
wave function the possibility of violation of quantum determinism
and quantum chaos initiation was shown.

%%%%%%%%%%%%%%%%%%%%%%%%%%%%%%%%%%%%%%%%%%%%%%%%%%%%%%%%%%%%%%%%%%%%%%%%%%
%%%%%%%%%%%%%%%%%%%%%%%%%%%%%%%%%%%%%%%%%%%%%%%%%%%%%%%%%%%%%%%%%%%%%%%%%%
%%                                                                      %%
%%                                                                      %%
%%   Глава 5. Заключение                              	                %%
%%                                                                      %%
%%                                                                      %%
%%%%%%%%%%%%%%%%%%%%%%%%%%%%%%%%%%%%%%%%%%%%%%%%%%%%%%%%%%%%%%%%%%%%%%%%%%
%%%%%%%%%%%%%%%%%%%%%%%%%%%%%%%%%%%%%%%%%%%%%%%%%%%%%%%%%%%%%%%%%%%%%%%%%%

\section{Conclusion}

In this work it was shown that the representation developed by the authors
includes not only Plank's constant $\hbar $, but new energy parameter as
well. Thus, when the energy of the particles collision exceeds a certain
critical value (which is different for the different systems), solution for
internal time $\tau $ coincides with an ordinary time - natural parameter  $s$.
In this case, the evolution equation for the system of bodies
transforms to ordinary nonstationary Schr\"odinger's equation. The scattering
process is in fact single-channel for this case.

But everything is quite different when the collision occurs below the
critical energy. As it is shown, in such a case the solution for
internal time $\tau $ in a definite range of $s$ has an oscillatory
character.  Moreover, for all the extreme points the derivative of $\tau $
with respect to $s$ has a jump of the first kind, while the phase portrait
of reactive (parametrical) oscillator has bifurcations. Let us note that these
are the collision modes that increase the interference effects, i.e. the
problem becomes essentially multichannel and includes the phase of resonant
state formation. At a small decrease of collisions energy, a number of internal
time oscillations grows dramatically. In this case the system loses all the
information about its initial state completely. Chaos arises in a wave
function, which then self-organizes into a new order within the limit $\tau
\rightarrow \infty $. Mathematically it becomes possible as a result of
common wave equation irreversibility by time (natural parameter $s$).

Let us stress that the result obtained supports the transitional complex
theory, developed by Eyring and Polanyi in the thirties \cite{Nikitin}
on the basis of heuristic considerations, the essence of which is statistical
description of chemical reactions. The amplitude of rearrangement transition
in three-body system is investigated in this work on the example of
$Li+\left( FH\right) _n\rightarrow \left( LiFH\right) ^{*}\rightarrow
(LiF)_m+H$ reaction and it is shown, that in the area where the quantity of
internal time peculiarities is high, it has a stochastic nature. It is also
been shown that the representation developed here satisfies the limit conditions
in the specified areas, including transition from $Q_{ch}$ area into $P$ area.
The latter occurs under $\hbar \rightarrow 0$ and at $E^i_k<E_c$, where $E_c$
is critical energy and $E^i_k$ is a collision energy.

\newpage

\begin{figure}
\setlength{\unitlength}{1 cm}
\begin{picture}(8,10.5)(-0.5,-10.5)
\put(0,0){\special {em:graph pic_1.gif}}
\end{picture}
\caption{Geodesic trajectories and dependence of internal time over natural
parameter for a) direct rearrangement reaction, b) direct reflection reaction
and c) rearrangement reaction going through resonance state.}
\end{figure}

\begin{figure}
\setlength{\unitlength}{1 cm}
\begin{picture}(8,5)(-5,-5)
\put(0,0){\special {em:graph pic_3.gif}}
\end{picture}
\caption{The regular map of collision energy $E^i_k$ and $x^2_0$ coordinate
initial values for passing (white fields) and reflecting (black fields)
geidesic trajectories.}
\end{figure}

\begin{figure}
\setlength{\unitlength}{1 cm}
\begin{picture}(8,7)(0,-7)
\put(0,0){\special {em:graph ready.gif}}
\end{picture}
\caption{The irregular map of collision energy $E^i_k$ and $x^2_0$ coordinate
initial values for passing (white fields) and reflecting (black fields)
geidesic trajectories. One can see that there is the selfsimilarity relative
to scale transformation in chaotic field.}
\end{figure}

\end{document}